\documentstyle[aps,prl,psfrag,epsf]{revtex}

\newcommand{\be}{\begin{equation}}
\newcommand{\ee}{\end{equation}}
\newcommand{\bea}{\begin{eqnarray}}
\newcommand{\eea}{\end{eqnarray}}

\newcommand{\ak}{a^{\dagger}}
\newcommand{\ket}[1]{| #1 \rangle}
\newcommand{\bra}[1]{\langle #1 |}

\begin{document}
\def\overlay#1#2{\setbox0=\hbox{#1}\setbox1=\hbox to \wd0{\hss #2\hss}#1
-2\wd0\copy1}
\twocolumn[\hsize\textwidth\columnwidth\hsize\csname@twocolumnfalse\endcsname

\title{Beyond single-photon localization at the edge of a Photonic Band Gap.}
\author{Georgios M. Nikolopoulos$^{1}$ and P. Lambropoulos$^{1,2}$}

\address{
  1. Department of Physics, University of Crete, Crete, Greece;\\
     and Institute of Electronic Structure {\rm \&} Laser, FORTH,	
     P.O. Box 1527, Heraklion 71110, Crete, Greece\\
  2. Max-Planck-Institut f\"ur Quantenoptik,
     Hans-Kopfermann-Str.\ 1, 85748 Garching, Germany}

\date{\today}
\maketitle
\begin{abstract}
We study spontaneous emission in an atomic ladder system, with both 
transitions coupled near-resonantly to the edge of a photonic band gap
continuum. The problem is solved through a recently developed technique and 
leads to the formation of a ``two-photon+atom'' bound state with fractional 
population trapping in both upper states. 
In the long-time limit, the atom can be found excited in a superposition of 
the upper states and a ``direct'' two-photon process coexists with the 
stepwise one. The sensitivity of the effect to the particular form of the 
density of states is also explored.
\end{abstract}
\vskip2pc]

\tighten

\section{INTRODUCTION}
The emergence of materials with photonic band gaps (PBG) 
\cite{yablo87,johnprl87}, also referred to as 
photonic crystals, has stimulated a broad range of problems pertaining to the 
interaction of few-level atoms with unusual (structured) reservoirs. The 
unconventional photonic density of states (DOS) associated with such materials 
leads to a number of novel effects which pose severe demands on the 
theoretical tools necessary for their prediction and/or interpretation. 
Certain basic issues related to such problems are encapsulated in the 
behavior of excited atomic states with transition frequencies around the 
edge of the DOS. Even the decay of an excited two-level atom (TLA) has been 
shown to exhibit novel behavior under such circumstances, the most notable 
aspect of which is the formation of the so-called ``photon+atom'' bound 
state (PABS) \cite{johnsd,kurizkisd,baytaipra}. Even in the context of this 
simple arrangement, however, it has 
been difficult to handle the problem of more than one photon in the PBG 
reservoir. An approach capable of surmounting this limitation has been 
presented recently in the context of a TLA and shown to provide the solution 
to the problem of the coupling of a TLA to both a PBG reservoir and a 
defect mode \cite{nikolg}.\\
For a number of reasons, three-level atoms are of particular interest in 
quantum optics and predictably their behavior in the context of structured 
reservoirs has been addressed. This includes ladder, lambda and V-type 
arrangements \cite{bayladder,baylambdaprl,zhu,tq,pasp}. 
Mathematical difficulties, however, have limited the relevant investigations 
to the case of only one photon present in the structured reservoir. This is 
indeed rather limiting as to the scope of questions one is allowed to 
contemplate.\\
It is our purpose in this paper to show how our approach provides a way out of 
this limitation. Any of the above mentioned standard three-level systems can 
be addressed. We have chosen the present results on the ladder system which 
in open space involves a cascade of two photon emissions \cite{tanscul}. 
As we show in the 
following sections, allowing both photons to be strongly coupled to the PBG 
reservoir we obtain what should be called ``two-photon+atom'' bound state. 
The formation of such a state is then found to be associated with a 
counterintuitive  coherent evolution of the three atomic states.\\
The first part of our study has been based on the so-called isotropic
model for the DOS of the PBG reservoir \cite{johnprb}, which could be argued 
to effectively reduce the problem to one dimension in wavevector 
$(\vec{k})$-space. 
A rather conspicuous feature of that DOS is a sharp (divergent) peak at the 
edge of the gap. In order to explore the persistence of our predictions under 
more relaxed assumptions for the DOS, we have introduced a modified form 
represented by a generalized Lorentzian profile.  
It is still isotropic, but it does not exhibit the singularity at the edge.  
The basic features of the predicted behaviour are still present with only 
some quantitative modifications.\\
The paper is organized as follows: Section II contains a description of 
the system and a brief summary of the discretization method.
In section III, we derive the equations of motion for the amplitudes involved
in the wavefunction of the system Atom+Continuum, which are solved
numerically. The dynamical behaviour of the system is thus investigated.  
In section IV the same problem is placed in the context of a generalized 
Lorentzian profile model for the DOS, while the results are summarized in 
section V. 
 
\section{THE SYSTEM}
\begin{figure}
  \begin{center}
    \leavevmode
    \epsfxsize8.5cm
    \epsfbox{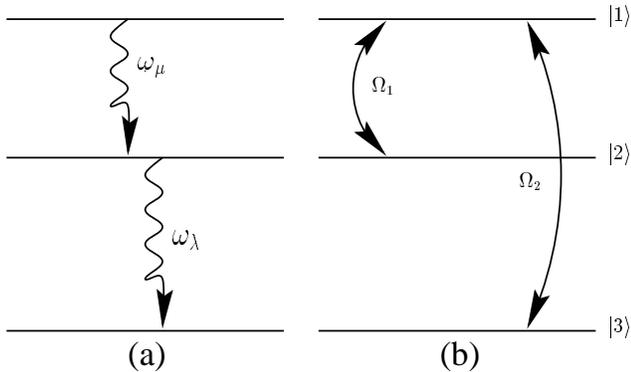}
  \end{center}
\caption{(a) Schematic representation of the atomic system and the possible
	transitions. (b) The upper state simultaneously coupled to the 
	intermediate state and the ground state via a single- and two-photon 
	process respectively.}
\label{system.fig}
\end{figure}

We consider a three level atom in a cascade configuration, with atomic levels 
$\ket{1}, \ket{2}, \ket{3}$ and energies $\hbar \omega_1, \hbar \omega_2, 0$ 
respectively, where $(\omega_1>\omega_2)$ (Fig. \ref{system.fig}a). 
Both atomic transitions are coupled 
near-resonantly to the edge of a photonic band gap and are thus strongly 
modified. In the isotropic model and close to the band-edge, the dispersion 
relation of PBG materials, can be approximated by the effective mass 
dispersion relation
\cite{johnprb,lew}
\be
\omega_k\simeq\omega_e+A(k-k_0)^2,
\label{dr}
\ee
where $\omega_e, k_0$ are the frequency and the wavenumber corresponding to 
the band-edge and A is a material specific constant. 
The corresponding density of states (DOS) reads
\be
\rho(\omega)=\frac{\rho_o}{\sqrt{\omega-\omega_e}}\Theta(\omega-\omega_e),
\label{dos}
\ee
where $\rho_o$ is a material specific constant and $\Theta(x)$ the 
Heaviside step-function.\\
Neglecting the zero-point energies of the field modes and adopting the 
rotating wave approximation, the Hamiltonian of the system in the interaction 
picture $(\hbar=1)$ is written as
\bea
V=i\sum_{\mu} g_{\mu}^{(1)}(\ak_{\mu}\sigma_{21} e^{-i\delta_{\mu}^1 t}
-a_{\mu}\sigma_{12} e^{i\delta_{\mu}^1 t})\nonumber\\
+i\sum_{\lambda} g_{\lambda}^{(2)}(\ak_{\lambda}\sigma_{32} e^{-i\delta_{\lambda}^2 t}-a_{\lambda}\sigma_{23} e^{i\delta_{\lambda}^2 t}),
\label{hamilt}
\eea
where $\sigma_{kl}$ denote the atomic dyadic operators $\ket{k}\bra{l}$ with 
$k,l\epsilon\{1,2,3\}$; $\delta_{\mu}^1=(\omega_1-\omega_2)-\omega_{\mu}$, 
$\delta_{\lambda}^2=(\omega_2-\omega_3)-\omega_{\lambda}$, 
while $a_{\mu}, \ak_{\mu} (a_{\lambda},\ak_{\lambda})$ are the creation and 
annihilation operators of the structured continuum, which is coupled to 
the atomic transitions, via the respective coupling constants 
$g_{\mu}^{(1)},g_{\lambda}^{(2)}$.\\ 
The spectral response $SR(\omega_\mu)$ corresponding to the PBG effective mass
dispersion relation Eq. (\ref{dr}) is given by
\be
SR(\omega_\mu)\equiv\sum_{\sigma}\int_{0}^{4\pi} d\Omega_\mu\rho(\omega_\mu)\left|g_\mu^{(1)}\right|^2=
\frac{{\cal C}_1}{\pi}\frac{\Theta(\omega_\mu-\omega_e)}{\sqrt{\omega_\mu-\omega_e}},
\label{SR1}
\ee
where the sum is over the polarizations, the integral runs over the solid 
angle and ${\cal C}_1$ is the effective coupling of the upper transition to 
the structured continuum \cite{bayladder}.\\ 
In order to deal with the double excitation into the structured 
reservoir, we use the discretization method that we have 
presented recently \cite{nikolg}. 
Briefly, we replace the density of modes in Eq. (\ref{dos}) near the 
atomic transitions (for $\omega<\omega_u$) by a collection of discrete 
harmonic oscillators, while the rest of the mode-density is treated 
perturbatively since it is far from resonance. We propagate the total 
wavefunction (Atom+Modes) to obtain the dynamics of the system. \\ 
For an arbitrary DOS, the frequencies of the discrete modes, are obtained 
through the relation
\be
\omega_{j+1}=\omega_{j-1}+2/\rho(\omega_j),
\label{disc1}
\ee
which for the case of the DOS of Eq. (\ref{dos}) can be reduced to
\be
\omega_j=\omega_e+j^2\delta\omega,
\ee
where $\delta\omega$ is chosen sufficiently small 
$(\delta\omega\approx4.4\times 10^{-4}{\cal C}_1^{2/3})$. 
The coupling ${\cal G}^{(1)}$ of the upper transition to each one of the 
discrete modes is found using Eq. (\ref{SR1})
\be
{\cal G}^{(1)}\approx\sqrt{\frac{{2\cal C}_1}{N\pi}{\sqrt{\omega_u-\omega_e}}},
\label{coup1}
\ee
where N is the number of discrete modes and $\omega_u$ is the 
upper limit of the discretized part. A similar relation can be derived
for the coupling ${\cal G}^{(2)}$ of the lower transition, where now 
${\cal C}_2$ will be the corresponding effective coupling.
In general, the dipole moments $\bra{1}d\ket{2}$ and $\bra{2}d\ket{3}$ are
different and thus for the rest of this paper we let the couplings 
${\cal C}_1$ and ${\cal C}_2$ be different.\\
Let us denote by $\ket{3, 1_j, 1_m}$, a state of the combined system 
(atom+field), where the atom is in state $\ket{3}$ and two photons have been 
emitted into the structured reservoir, populating the modes $j$ and $m$ 
respectively. Following this pattern of notation, the wavefunction of the 
complete system reads
\bea
\ket{\Psi(t)}=a\ket{1, 0, 0}+\sum_j b_j\ket{2, 1_j, 0}+
\sum_{j,m}C_{jm}\ket{3, 1_j, 1_m}.
\eea

\section{EQUATIONS OF MOTION AND RESULTS}
\begin{figure}
  \begin{center}
    \leavevmode
    \epsfxsize8.5cm
    \epsfbox{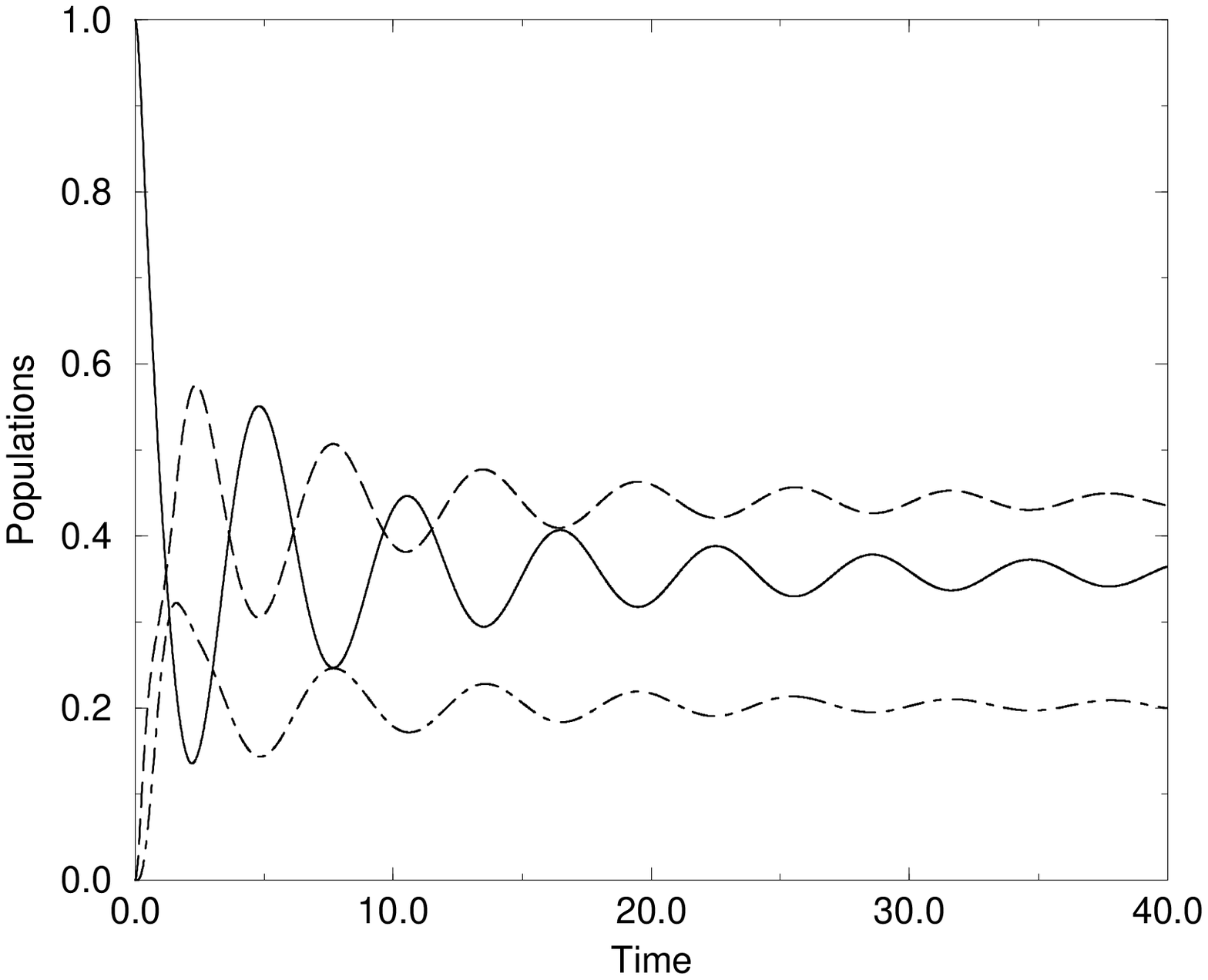}
  \end{center}
\caption{The population in the atomic states as function of time. The
    	solid line is for $\ket{1}$, the dashed line for $\ket{2}$ and 
    	the dot-dashed line for $\ket{3}$.
    	Parameters: ${\cal C}_2=1.5{\cal C}_1$, 
	$\delta_{12}=-{\cal C}_2^{2/3}$ 
    	and $\delta_{23}=0$. The time is in units of ${\cal C}_1^{2/3}$.}
\label{plot1.fig}
\end{figure}

\begin{figure}
  \begin{center}
    \leavevmode
    \epsfxsize8.5cm
    \epsfbox{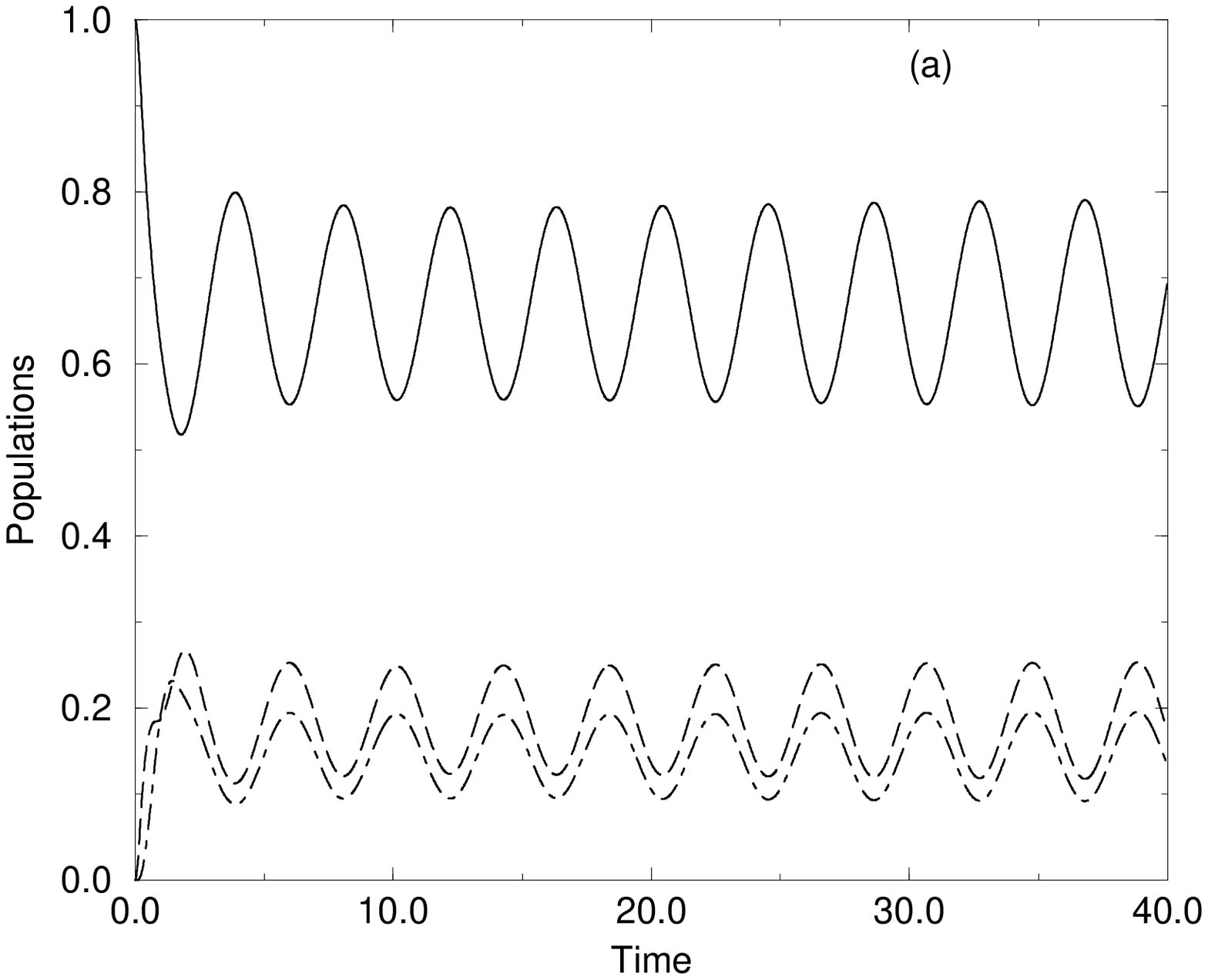}
  \end{center}
  \begin{center}
    \leavevmode
    \epsfxsize8.5cm
    \epsfbox{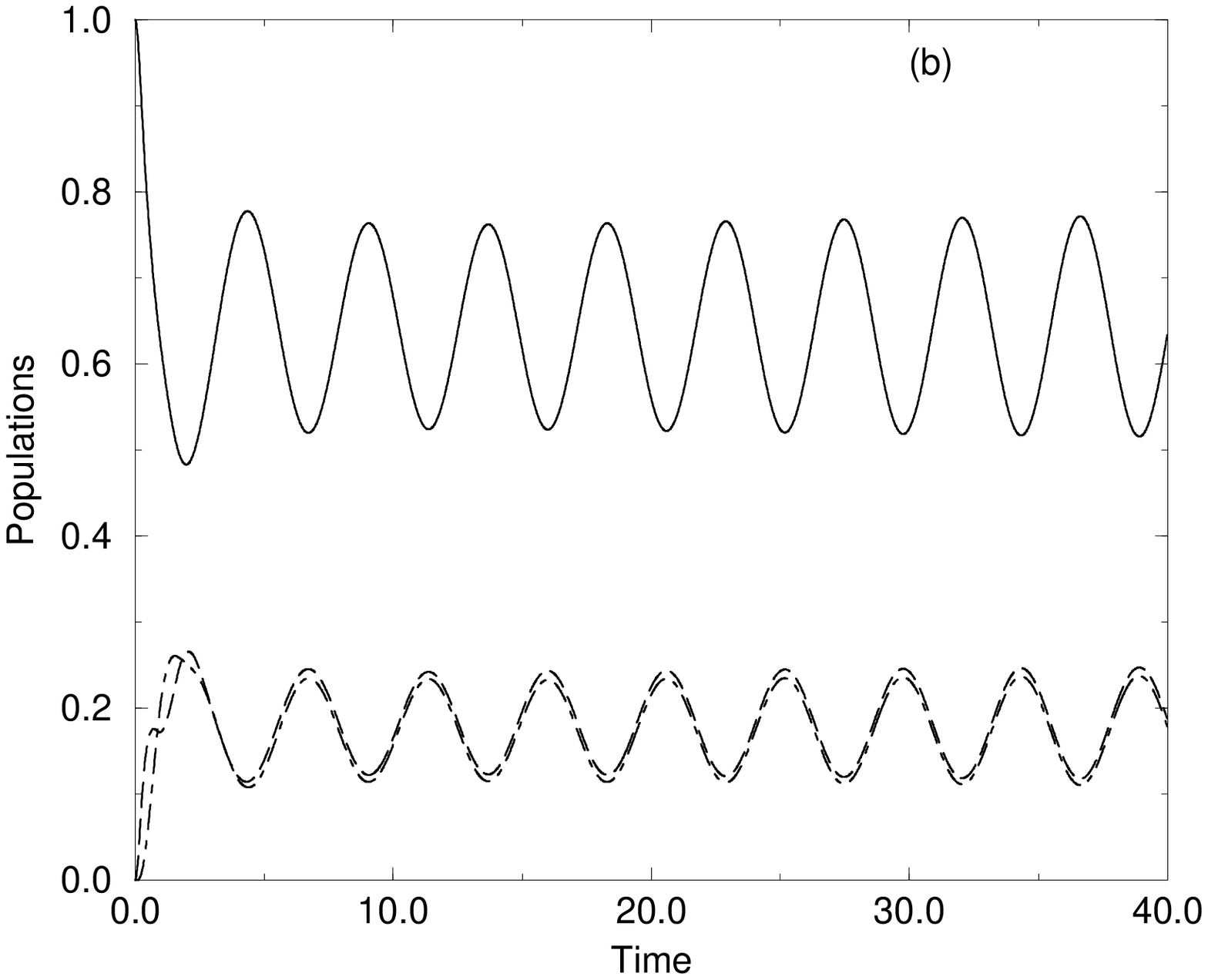}
  \end{center}
\caption{The population in the atomic states as function of time. The
    solid line is for $\ket{1}$, the dashed line for $\ket{2}$ and 
    the dot-dashed line for $\ket{3}$.
    Parameters: ${\cal C}_2=1.5{\cal C}_1$, 
    (a) $\delta_{12}=-2{\cal C}_2^{2/3}$ and $\delta_{23}=1{\cal C}_1^{2/3}$; 
    (b) $\delta_{12}=-2{\cal C}_2^{2/3}$ and $\delta_{23}=1{\cal C}_2^{2/3}$. 
    The time is in units of ${\cal C}_1^{2/3}$.}
\label{plot2.fig}
\end{figure}

\begin{figure}
  \begin{center}
    \leavevmode
    \epsfxsize8.5cm
    \epsfbox{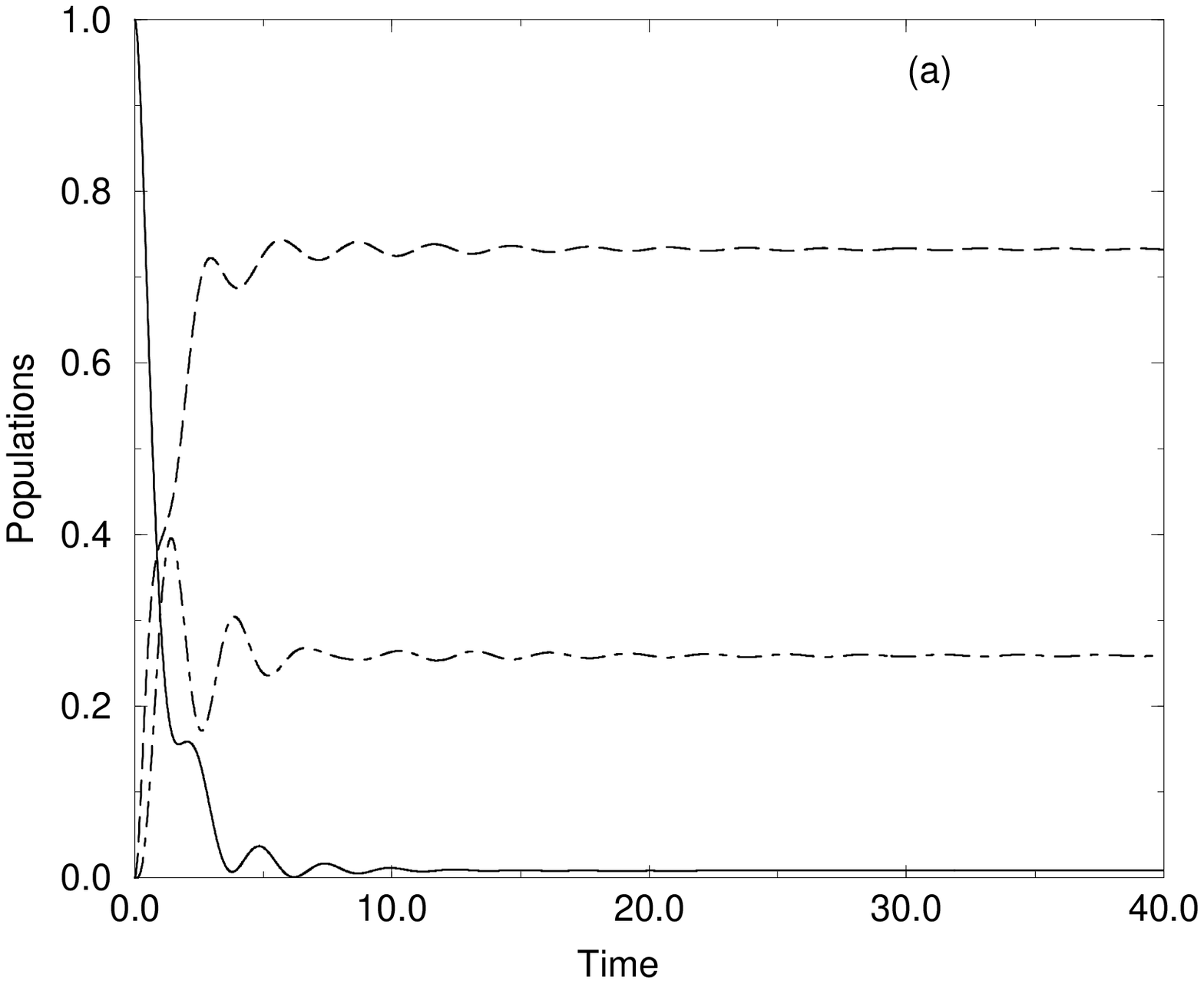}
  \end{center}
  \begin{center}
    \leavevmode
    \epsfxsize8.5cm
    \epsfbox{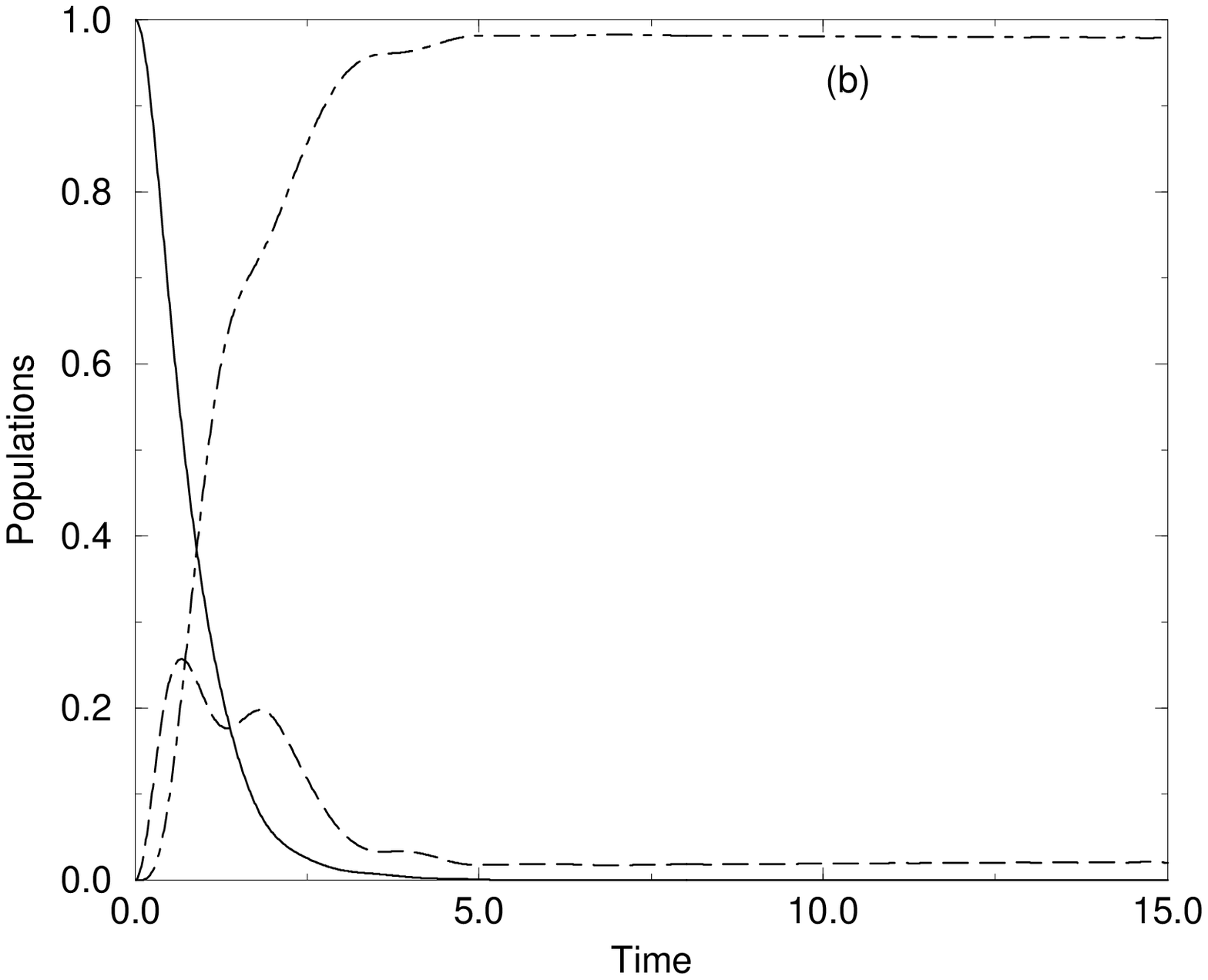}
  \end{center}
\caption{The population in the atomic states as function of time. The
    solid line is for $\ket{1}$, the dashed line for $\ket{2}$ and 
    the dot-dashed line for $\ket{3}$.
    Parameters: ${\cal C}_2=1.5{\cal C}_1$, 
    (a) $\delta_{12}=1{\cal C}_2^{2/3}$ and $\delta_{23}=-1{\cal C}_1^{2/3}$;
    (b) $\delta_{12}=2{\cal C}_2^{2/3}$ and $\delta_{23}=3{\cal C}_1^{2/3}$.
    The time is in units of ${\cal C}_1^{2/3}$.}
\label{plot3.fig}
\end{figure}

\begin{figure}
  \begin{center}
    \leavevmode
    \epsfxsize8.5cm
    \epsfbox{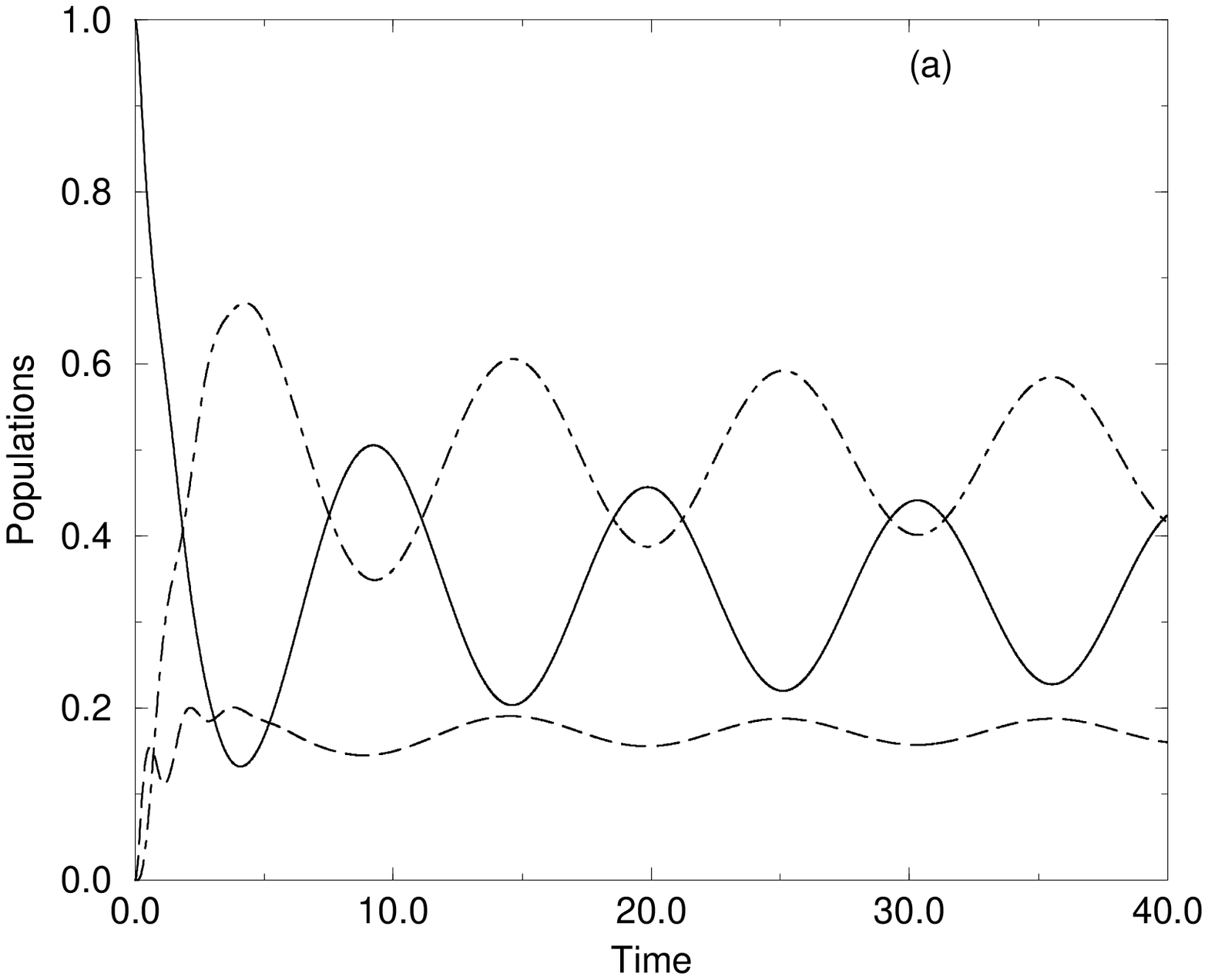}
  \end{center}
  \begin{center}
    \leavevmode
    \epsfxsize8.5cm
    \epsfbox{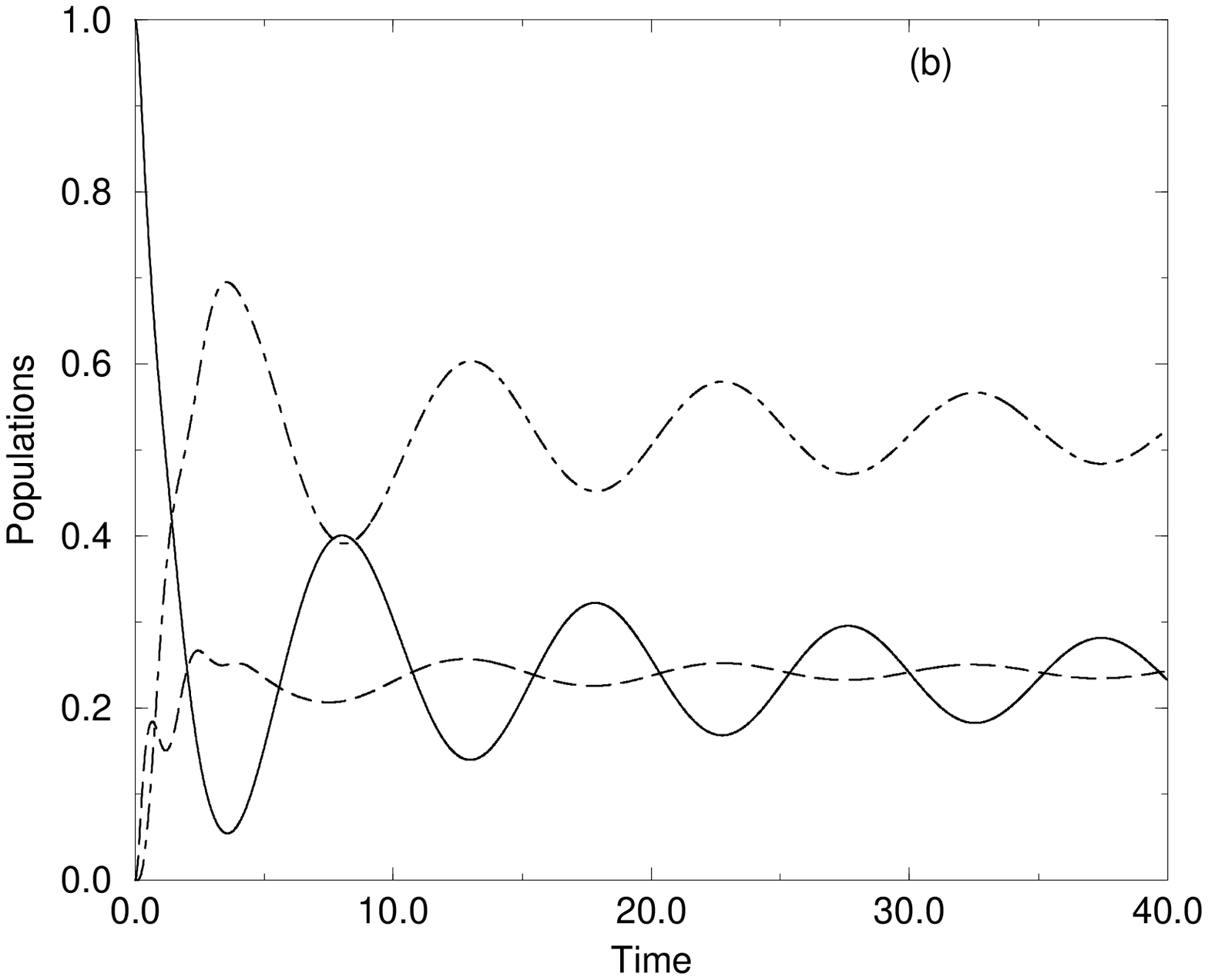}
  \end{center}
\caption{The population in the atomic states as function of time. The
    solid line is for $\ket{1}$, the dashed line for $\ket{2}$ and 
    the dot-dashed line for $\ket{3}$.
    Parameters:  ${\cal C}_2=1.5{\cal C}_1$, 
    (a) $\delta_{12}=-2{\cal C}_2^{2/3}$ and $\delta_{23}=2{\cal C}_2^{2/3}$; 
    (b) $\delta_{12}=-2{\cal C}_1^{2/3}$ and $\delta_{23}=2{\cal C}_1^{2/3}$.
    The time is in units of ${\cal C}_1^{2/3}$.}
\label{plot4.fig}
\end{figure}

\begin{figure}
  \begin{center}
    \leavevmode
    \epsfxsize8.5cm
    \epsfbox{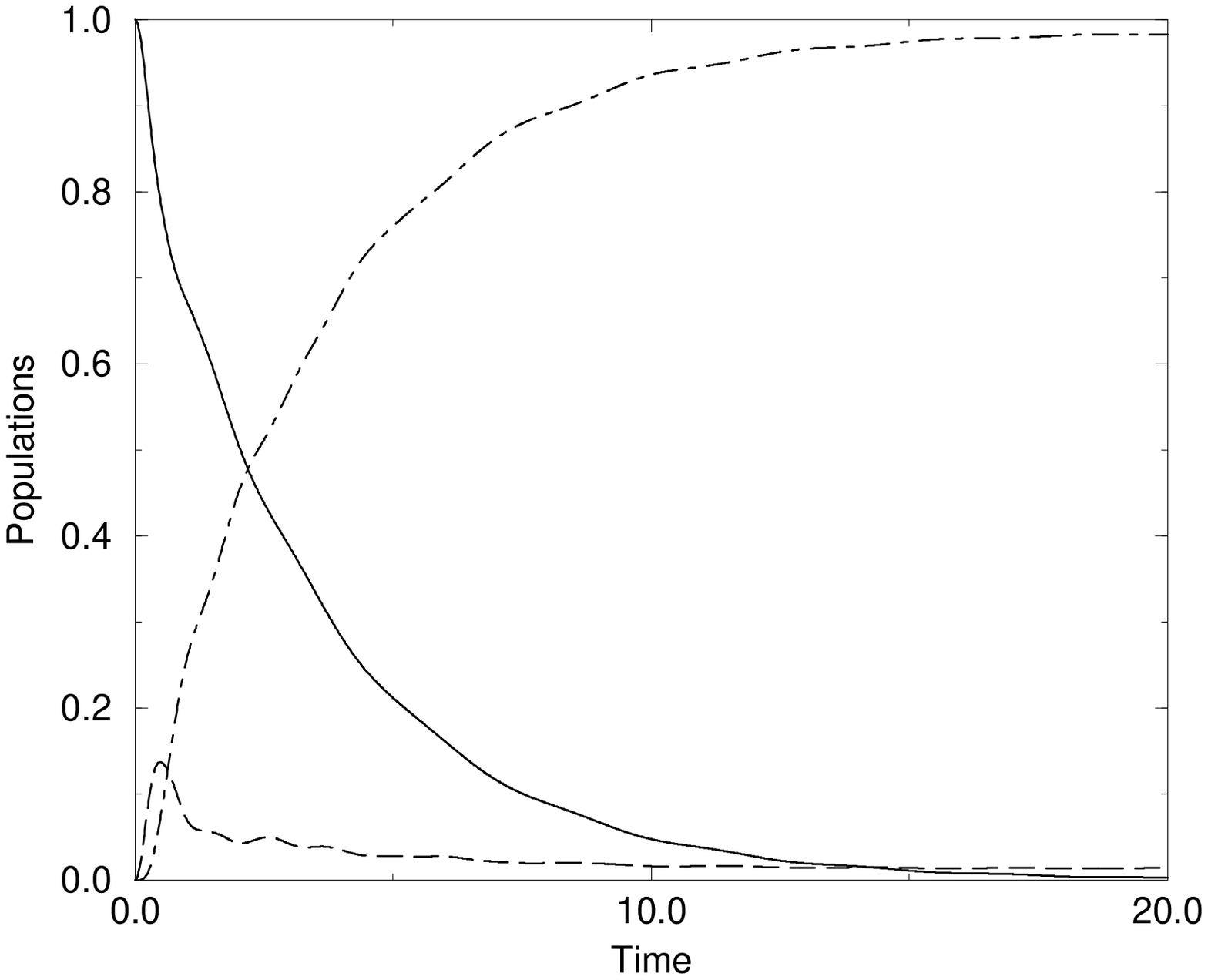}
  \end{center}
\caption{The population in the atomic states as function of time. 
    The solid line is for $\ket{1}$, the dashed line for $\ket{2}$ and 
    the dot-dashed line for $\ket{3}$.
    Parameters: ${\cal C}_2=1.5{\cal C}_1$, 
    $\delta_{12}=-2{\cal C}_2^{2/3}$ and $\delta_{23}=4{\cal C}_2^{2/3}$.
    The time is in units of ${\cal C}_1^{2/3}$.} 
\label{plot5.fig}
\end{figure}

The time dependence of the amplitudes, is governed by the Schr\"odinger equation
and after eliminating the off-resonant modes \cite{nikolg}, we find
\bea
\dot{a}&=&-i\sum_{\omega_{\mu}>\omega_u}\frac{\left|g_\mu^{(1)}\right|^2}{\delta_{\mu}^1}a-
\sum_{j=1}^N {\cal G}_{j}^{(1)} b_j e^{i\delta_j^1 t},
\label{ampl_eq1}
\\
\dot{b_j}&=&-i\sum_{\omega_{\lambda}>\omega_u}\frac{\left|g_\lambda^{(2)}\right|^2}{\delta_{\lambda}^2}b_j
-\sum_{m=1}^N {\cal G}_{m}^{(2)} C_{jm} e^{i\delta_m^2 t}\nonumber \\
&&-\sqrt{2} {\cal G}_{j}^{(2)} C_{jj} e^{i\delta_j^2 t}+
{\cal G}_{j}^{(1)} a e^{-i\delta_j^1 t},
\\
\dot{C_{jm}}&=&{\cal G}_m^{(2)} b_j e^{-i\delta_m^2 t}+
{\cal G}_j^{(2)} b_m e^{-i\delta_j^2 t},
\\
\dot{C_{jj}}&=&\sqrt{2} {\cal G}_j^{(2)} b_j e^{-i\delta_j^2 t},
\label{ampl_eq2}
\eea
where $j, m$ are mode indices and for all discrete modes 
${\cal G}^{(1)}={\cal G}_j^{(1)}={\cal G}_m^{(1)}$ and 
${\cal G}^{(2)}={\cal G}_j^{(2)}={\cal G}_m^{(2)}$, while the shift terms can 
be determined, converting the sum into an integral from $\omega=\omega_u$ to 
infinity.\\
This set of equations is solved numerically for 150 discrete modes 
$(\omega_u\approx10{\cal C}_1^{2/3})$, with the 
results presented in Figs. \ref{plot1.fig}-\ref{plot5.fig}. 
We plot the population in the upper level (solid line), the intermediate level
(dashed line) and the lower level (dot-dashed line), for various detunings of 
the transition frequencies from the band-edge. 
We define the relative detunings of the upper and lower transitions from the 
band-edge as  $\delta_{12}=\omega_1-\omega_2-\omega_e$ and 
$\delta_{23}=\omega_2-\omega_3-\omega_e$, respectively.\\
 In all figures, we can identify a ``transient regime'', on a short time 
scale of the order of ${\cal C}_1^{2/3}$, when part of the atomic 
population is lost. On a longer time scale 
(``dynamic regime''), the populations in the atomic 
levels undergo oscillations, strongly dependent on the relative detunings
from the band-edge, reflecting the emission and reabsorption of the photon(s).
The localization at the atom of the photon emitted in the first transition, 
is accompanied by emission of a photon in the lower transition, which will be 
also localized at the atom. In analogy with the case of one photon in a TLA, 
two photons are now backscattered to the atom after tunneling a characteristic 
distance and reexcite it.\\
As is known from the coupling of a TLA to the PBG reservoir, the dressing of 
the atom by its own radiation causes splitting of the atomic levels. 
This splitting is sufficiently strong to
push one level of the doublets outside the gap and the other inside.
The dressed state outside the gap looses all its population in the long-time 
limit, while the one inside the gap is protected from dissipation and thus is
stable. The number of stable localized states, is intimately connected to the
behavior of the system in the long-time limit. One such state gives rise to 
steady state population in the excited level, while two of them lead to an 
undamped beating of the system between the two non-decaying states 
\cite{kurizkisd,baytaipra}.\\ 
For our three-level atom, two transitions are at the edge, and thus more than 
one stable localized state can be found in the gap. 
In analogy to the ``single-photon+atom'' bound state, we have the formation 
of a ``two-photon+atom'' bound state, which exhibits population trapping in 
both excited states, in the long-time limit. Thus, the atom is finally 
excited, in a superposition of the upper states (Fig. \ref{plot1.fig}).
This is a novel behavior, due to the fact that both transitions and not only 
one, are coupled to the same structured continuum. Note that, even in the case
that only the $(\ket{2}\rightarrow\ket{3})$ transition is at the edge of the 
gap, it is the intermediate level that exhibits non-zero steady-state 
population but not the upper level \cite{johnsd,bayladder}.\\
In the language of dressed states, the oscillations in the populations 
of the atomic levels reflect the interference between the dressed states of 
the atom. We additionally note that the oscillations in the populations of the 
intermediate and lower level are "in phase". 
This is a rather surprising result, since one would expect the population in 
the intermediate state to be maximum when the population in the lower state 
is minimum and vise-versa.
The phenomenon is even more pronounced in Figs. \ref{plot2.fig}. 
After an initial transient regime in which about $15\%$ of the population is 
lost, the remaining population oscillates between the upper level and the two 
lower levels, in a non-dissipative way. The undamped oscillations imply the 
beating of the system, between more than two non-decaying dressed states.\\
The ``in phase'' oscillations, stem from the localization of both photons at 
the site of the atom. The system can make the transitions 
$\ket{1} \leftrightarrow \ket{3}$ either with a stepwise process 
($\ket{1}\leftrightarrow\ket{2}\leftrightarrow\ket{3}$) or with a 
two-photon process ($\ket{1} \leftrightarrow \ket{3}$). 
Which of the two routes the system will follow to arrive at $\ket{3}$ depends
on the detuning $\delta_{12}$.\\
Specifically, if the upper transition is outside the gap, whatever 
the detuning from the edge of the lower transition is, it seems 
preferable for the system to "decay" via the stepwise process rather 
than the two-photon process indicated in Fig.4.  This is no different 
from the behaviour of a ladder system in open space. 
On the contrary, if the upper transition is inside the gap, the 
system evolves in time as if the upper state were coupled to the 
intermediate state via a single-photon process, and simultaneously to 
the ground state via a "direct" two-photon process, 
with respective frequencies $\Omega_1$ and $\Omega_2$.  
This can not be anticipated on the basis of the behaviour in open space and 
it is what we meant by counter-intuitive in the introduction.\\
To gain further insight into this effect, we can adopt a simple 3-level model 
without dissipation and assign a single-photon Rabi frequency $\Omega_1$  
between $\ket{1}$ and $\ket{2}$ and a two-photon Rabi frequency $\Omega_2$ 
between $\ket{1}$ and $\ket{3}$ (Fig. \ref{system.fig}b).
An analysis through standard rate equations shows that the populations of the 
atomic levels oscillate with the same frequency 
$\Omega=\sqrt{\Omega_1^2+\Omega_2^2}$, with the 
oscillation of the two lower levels being in phase.  The ratio 
of the amplitude of the oscillation of the intermediate level 
to that of the lower level is related to the ratio of the Rabi 
frequencies, i.e. ${\Omega_1}/{\Omega_2}$.  In Fig.  \ref{plot2.fig}b 
representing the result 
of the numerical calculation for the system in the PBG reservoir, 
we note that the dashed and the dot-dashed lines, corresponding 
to the populations in the intermediate and ground levels, respectively, 
are indistinguishable.  This implies that the corresponding effective 
Rabi frequencies $\Omega_1$ and $\Omega_2$ are practically equal.  
It is the combination of the coupling constants and detunings that conspire 
to produce that behaviour.\\ 
The effective detuning of the $\ket{3} \leftrightarrow \ket{1}$ 
transitions from the band-edge for the ``direct'' two-photon process is 
defined as $\Delta^{(2)}=\omega_1-\omega_3-2\omega_e$. 
From the known dynamics of a TLA with transition frequency at the edge of the 
gap, depending on $\Delta^{(2)}$, we may expect suppression or inhibition of 
the ``direct'' two-photon process. 
Specifically, for detunings inside the gap $(\Delta^{(2)}<0)$,
the ``direct'' two-photon emission should be totally or partially suppressed, 
while for detunings outside the gap and almost at the edge $(\Delta^{(2)}>0)$,
it should be enhanced due to the high density of final available states.\\ 
 For transitions symmetrically placed around the band-edge 
$(\delta_{12}=-\delta_{23})$, with the upper one being inside the 
gap $(\delta_{12}<0)$, the detuning for the ``direct'' two-photon transition 
is exactly at the edge $(\Delta^{(2)}=0)$ (Fig. \ref{plot4.fig}). 
For the parameters used in Figs. \ref{plot4.fig} and 
without taking into account the ``direct'' two-photon process, the 
``single-photon+atom'' bound state for the upper state should be metastable, 
in the sense that the main part of the population should be lost in the 
long-time limit. 
On the contrary, we find that this does not happen and the part 
of the population that has not been lost in the transient regime, oscillates 
between the upper and the ground state. 
These oscillations are not reflected in the intermediate 
level's population which remains almost constant with some oscillations of 
negligible amplitude. This behavior definitely indicates the coupling of the 
upper level to the ground level via a ``direct'' two-photon process as 
described above. 
Note again the ``in phase'' oscillations for the two lower levels.\\   
Choosing $\delta_{23}$ such that $\Delta^{(2)}>0$ 
(Fig. \ref{plot5.fig}), the main part of the population is indeed 
lost in the long-time limit.
The difference between Fig. \ref{plot4.fig}a and Fig. \ref{plot5.fig} is the 
detuning of the lower transition from the band-edge. 
In both cases ($\delta_{23}=2{\cal C}_2^{2/3}$, 
$\delta_{23}=4{\cal C}_2^{2/3}$), the behavior of a TLA in the 
long-time limit is the same i.e. the population is lost. For the Ladder 
system, however, we note an oscillatory behavior where part 
of the population is trapped to the atom in the long-time limit for 
$\delta_{23}=2{\cal C}_2^{2/3}$ and complete decay for 
$\delta_{23}=4{\cal C}_2^{2/3}$. 
The photon-atom bound state formed due to the upper transition becomes 
therefore metastable\cite{johnsd,bayladder} as soon as $\delta_{23}$ 
is chosen such that $\Delta^{(2)}>0$. 
This is a conclusion that has been checked for various detunings of 
the atomic transitions from the band-edge but with the upper one always in 
the gap ($\delta_{12}\leq0$).  

\section{Gap with a Lorentzian Profile of DOS}
\begin{figure}
  \begin{center}
    \leavevmode
    \epsfxsize8.5cm
    \epsfbox{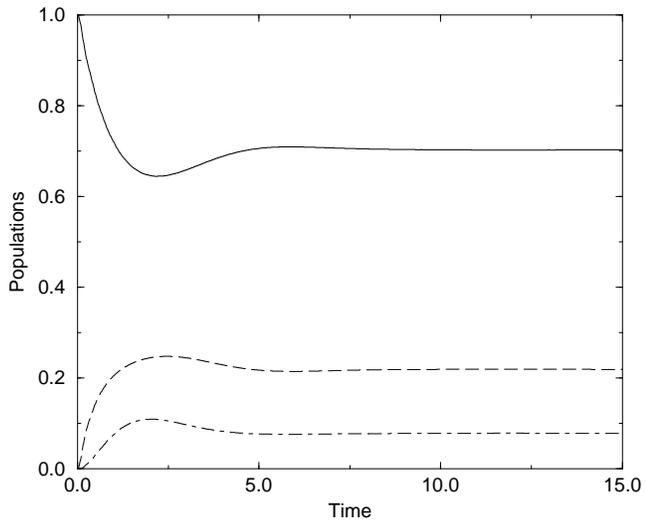}
  \end{center}
\caption{The population in the atomic states as function of time. 
    The solid line is for $\ket{1}$, the dashed line for $\ket{2}$ and 
    the dot-dashed line for $\ket{3}$.
    Parameters: $\gamma_1=0.5\gamma_2$, $\Gamma=\gamma_2$, 
    $\delta_{12}=0.1\gamma_2$, $\delta_{23}=0.3\gamma_2$ and 
    $\omega_{up}=-\omega_{low}=20\gamma_2$. 
    The time is in units of $\gamma_2$.}
\label{lor1.fig}
\end{figure}

The isotropic model for the DOS of a photonic crystal has been employed 
quite extensively in the literature. That is the model we have employed in 
the previous sections in order to have a direct assessment of the new results 
vis a vis those obtained in previous work with only one photon in the 
reservoir, where the same DOS has been assumed. 
It is, however, well known that this DOS 
is somewhat artificial, with a divergence as the frequency approaches the band 
edge. The infinitely high sharp peak at the edge tends to exaggerate many 
effects and it is important to check predictions made on the basis of this 
model against other forms of DOS. With this intention, we introduce here a DOS 
which although still isotropic, does not exhibit a divergence at the edge.\\
What is essential for an appropriate model of the DOS is that it exhibit a 
dip and also that it tend to the open space DOS as the frequency becomes much 
larger or smaller than the mid-gap frequency. We have chosen to adopt as a 
model of such a DOS an inverted Lorentzian of higher order given by the 
expression
\be
\rho(\omega)=\rho_0\left [1-
\frac{\Gamma^n}{(\omega-\omega_o)^n+\Gamma^n}\right ].
\label{rho}
\ee
First of all, this DOS approaches the open space value $\rho_0$ for 
$|\omega-\omega_o|>>\Gamma$. Second, it does not exhibit a divergence at the 
edge. In fact, the ``edge'' is not infinitely steep, but does rise more 
steeply, as $n\rightarrow\infty$.\\
It could be argued that this DOS does not exhibit a clear edge and possesses a 
zero only at one point. It should be kept in mind on the other hand that, in a 
realistic PBG material, the gap does not necessarily mean a true zero but a 
range of 
frequencies over which the DOS is several orders of magnitude smaller than 
that of open space \cite{dosref}. Taking $n$ sufficiently large, in Eq. 
(\ref{rho}), one can obtain a range of frequencies over which 
$\rho(\omega)/\rho_0$ is smaller than a desired value. For $n=6$, for example,
 $\rho(\omega)/\rho_0\leq 10^{-6}$ for 
$\omega=\omega_o\pm0.1\Gamma$. One can further combine the inverted Lorentzian 
with step functions in order to simulate a true zero over a range of $\omega$, 
if so desired. The shape of the DOS given by Eq. (\ref{rho}), can be viewed as 
a compromise between the isotropic and the anisotropic models \cite{johnprb}, 
and has been used in the literature for $n=2$ 
\cite{lew,nab,gar}. It should be stressed once more that the modified DOS 
in no way relaxes its isotropic in $\vec{k}$ space nature. It only eliminates 
the singularity.\\
We proceed now to the exploration of the ladder system under the DOS of 
Eq. (\ref{rho}) adopting the specific case of $n=6$. It is one of the 
strengths of the technique of discretization that it can be implemented with 
essentially any DOS, provided the manner of discretization is adapted to the 
demands of the particular form. 
For the case under consideration, which is symmetric around 
$\omega_o$, we chose a range $\omega_{low}<\omega<\omega_{up}$ within which 
the DOS is replaced by a sufficiently large number (in this case 150) 
equidistant discrete modes.   
The spectral response corresponding to Eq. (\ref{rho}) for $n=6$ is of the 
form 
\be
SR^{(1)}(\omega_\mu)=\frac{\gamma_1}{2\pi}\left [1-
\frac{\Gamma^6}{(\omega_{\mu}-\omega_o)^6+\Gamma^6}\right ],
\label{SR6}
\ee
where $\gamma_1$ is the decay rate of the upper state in free space.\\
The coupling of the upper transition to the $i^{th}$ discrete mode is 
frequency dependent and is given by
\be
{\cal G}^{(1)}(\omega_i)=\sqrt{\frac{\gamma_1}{2\pi}}\sqrt{\left [1-
\frac{\Gamma^6}{(\omega_i-\omega_o)^6+\Gamma^6}\right ]\Delta\omega},
\label{couplor}
\ee
where $\Delta\omega\approx 0.27\gamma_2$ 
is the spacing between two discrete modes and $\omega_i$ 
is the frequency corresponding to the $i^{th}$ mode. An analogous relation 
gives the coupling of the lower transition, ${\cal G}^{(2)}$, to the $i^{th}$ 
mode, where 
$\gamma_1$ is replaced by $\gamma_2$, the free space decay-rate of the 
intermediate state. By way of comparison illustrating the philosophy of the 
discretization procedure, note that here the mode-spacing is uniform but the 
coupling constant frequency-dependent; while in section II 
(Eqs. \ref{disc1} and \ref{coup1}) it is the other way around.\\ 
The rest of the mode-density, for $\omega>\omega_{up}$ and 
 $\omega<\omega_{low}$, is treated perturbatively \cite{nikolg} leading to a 
shift for the two upper levels. These shift terms differ from those 
in Eqs. (\ref{ampl_eq1}-\ref{ampl_eq2}) and are not given explicitly. They 
can, however, be determined numerically. The relative positions of the upper 
and lower transition frequencies are now defined with respect to the 
central frequency $\omega_o$ of the gap, in terms of the detunings: 
$\delta_{12}=\omega_1-\omega_2-\omega_o$ and 
$\delta_{23}=\omega_2-\omega_3-\omega_o$, respectively.  
In Fig. \ref{lor1.fig}, we present the evolution of the population in the 
states of the Ladder system as a function of time, for a particular 
combination of detunings. 
As in the previous figures corresponding to the isotropic 
model, both upper and lower levels exhibit non-zero steady-state population, 
as a consequence of the ``two-photon+atom'' bound state. The oscillations, 
however, in the atomic populations, which can be interpreted as interference 
between the dressed states, are not present. It seems that this oscillatory 
behavior is strongly related to the isotropic model. For the Lorentzian 
profile, the part of the doublet that is pushed outside the 
gap, decays much faster than the isotropic model would predict. From this we 
conclude that in the isotropic model, the ``dynamic regime'' which follows the 
initial transient regime and is dominated by the emission and reabsorption of 
photons, or else the interference between the various dressed states, is much 
more pronounced than in the Lorentzian model, or we would argue any model 
that does not exhibit the highly peaked feature of the isotropic model 
\cite{vats}. Without the concentration of the density of states around a peak, 
the notion of dressed states is diluted. 
Nevertheless, the coherent superposition of states and the 
``two-photon+atom'' bound state persists even in this model. 
Atomic populations remain trapped for long times, long compared to 
$1/\gamma_2$.\\
A further issue needs to be brought up here, in connection with the time scale 
of the persistence of any ``photon+atom'' bound state. As long as the model 
for the DOS involves an exact zero over some frequency range, the 
``photon+atom'' bound state lives for ever, which mathematically 
implies a non-zero 
fractional population trapping in the limit $t\rightarrow\infty$. In reality, 
however, the DOS, more often than not, will not involve an exact zero but a 
deep minimum, as already mentioned in the beginning of this section. 
As a result, the life-time of the ``photon+atom'' bound state, will be long 
on some time scale but not necessarily for $t\rightarrow\infty$ in the 
mathematical sense. It is this finiteness of the lifetime of the 
``two-photon+atom'' bound state that would cause the 
populations of states $\ket{1}$ and $\ket{2}$ in Fig.7 to decay to zero, if 
the calculation were extended to sufficiently long times. The essential point 
therefore is that the population may remain trapped for such a long time 
which, for all practical purposes, is equivalent to $t\rightarrow\infty$.

\section{SUMMARY}      
We have investigated the dynamics of a Ladder atomic system
with both transitions coupled to the same structured reservoir. This has been 
possible through a discretization approach according to which 
the density of modes over a range near the atomic transitions is replaced 
by an appropriately chosen collection of discrete 
harmonic oscillators, with the rest of the mode-density treated perturbatively.
We have found that this system supports a ``two-photon+atom'' bound state 
which leads to a fractional population trapping in both of the upper states 
and the atom can be in a superposition of the upper levels even in the 
long-time limit. 
In the presence of the two photons at the site of the atom, we have shown 
that the atom has two paths for the $\ket{1} \leftrightarrow \ket{3}$ 
transition, and found that a ``direct'' two-photon process coexists 
with a stepwise one. 
Which of the two dominates is determined mainly by the detuning 
of the upper transition from the band-edge.
We have further explored the persistence of this effect under much more 
relaxed forms of the DOS and shown that, although quantitatively modified, 
the basic effect remains.

\bibliographystyle{prsty}

\newpage

\newpage

\newpage

\newpage

\newpage

\newpage

\end{document}